\documentclass[sigconf, nonacm]{acmart}

\usepackage{pvldb}
\usepackage{macro}
\usepackage[inline]{enumitem}
\usepackage{tabularx}

\definecolor{darkgreen}{rgb}{0.0, 0.5, 0.0}

\renewcommand\vldbdoi{XX.XX/XXX.XX}
\renewcommand\vldbpages{XXX-XXX}
\renewcommand\vldbavailabilityurl{URL_TO_YOUR_ARTIFACTS} 

\begin{document}
\title{TrajectoryDB: A New Database for Agent Trajectories [Vision]}

\author{Yunjia Zheng}
\affiliation{%
  \institution{Harvard University}
  \city{Cambridge}
  \country{United States}
}
\email{yunjia\_zheng@g.harvard.edu}

\author{Juncheng Yang}
\affiliation{%
  \institution{Harvard University}
  \city{Cambridge}
  \country{United States}
}
\email{juncheng@seas.harvard.edu}

\begin{abstract}
AI agents generate rich execution trajectories that capture their interactions with large language models, tools, and external environments. These trajectories are increasingly valuable for downstream tasks such as memory extraction, model fine-tuning, runtime optimization, and security and cost monitoring. Yet trajectory data today is fragmented across files, databases, and observability systems, with no persistent data management system designed around its unique structure and access patterns.

We argue that trajectories should be treated as a distinct data type. A trajectory combines hierarchical execution structure, large volumes of text whose analysis often requires semantic reasoning, and rich dependencies and lineage among events, intermediate states, and derived artifacts. These properties introduce new requirements throughout the data lifecycle. Ingestion must reconstruct and preserve execution structure and lineage; storage must efficiently organize large but highly redundant contexts while maintaining relationships among records; and query processing must jointly reason over structure, temporal order, semantics, and lineage.

We therefore envision \textbf{TrajectoryDB}, a trajectory-native data management system that co-designs ingestion, storage, and query processing to efficiently manage and analyze agent execution trajectories.
\end{abstract}

\maketitle

\vldbtopmatter

\section{Introduction}

Large language models (LLMs) generate text from an input sequence. An \emph{agent} places an LLM in a runtime loop: when the model response contains a tool call, the runtime executes it in the environment and feeds the result into the next model invocation~\cite{yao2023react}. This loop turns model outputs into actions and allows tasks to unfold over many rounds of reasoning, interaction, and execution. Agents are now used for repository-level software engineering~\cite{yang2024sweagent}, end-to-end data analysis~\cite{hong2025datainterpreter}, and customer-service workflows combining user interaction with domain-specific APIs~\cite{yao2025taubench}. Their adoption is also moving rapidly into production. 
On OpenRouter, agent-driven token consumption grew roughly $14\times$ from February to August 2026 and reached nearly $5\times$ that of human-driven usage~\cite{a16z2026agentictokens}.

As agents interact with models and their environments, they generate a rich execution history, or \emph{trajectory}. A trajectory records the messages sent to the LLM, model responses, tool calls, tool results, and associated runtime state. These trajectories are increasingly treated as persistent datasets rather than ephemeral logs. They can be mined for reusable memories that guide future behavior, reused as training data for model fine-tuning, and analyzed for failure diagnosis and safety control~\cite{deshpande2025trail,barke2026agentrx,openai2026longhorizon}. By connecting agent behavior with the underlying runtime, trajectories also support performance diagnosis, cost attribution, and resource-aware scheduling~\cite{lin2024parrot,luo2026agentix,wang2026mars}.

Yet existing systems manage only fragments of the trajectory lifecycle. Research systems commonly store trajectories as JSONL files and process them using task-specific Python pipelines~\cite{zhao2024expel,fu2024autoguide,ouyang2026reasoningbank,zhang2026ace,yang2025swesmith}. Such pipelines preserve the raw trajectory but provide little support for persistent, reusable derived state or declarative queries. Production \emph{memory} systems such as Mem0~\cite{chhikara2025mem0} and Letta~\cite{letta2026memory} take the opposite approach: they persist extracted memories using structured storage, similarity search, and graph retrieval, but typically expose only task-specific derived state rather than the complete relationship between that state and its source trajectory. Security and observability systems ingest agent activity into mature log stores~\cite{opentelemetry2026logs}, but their data models are designed around individual events. Reconstructing the context visible to a model invocation, traversing a subtrajectory, or tracing a semantic judgment back to the exact messages and tool results that support it therefore requires application-specific processing. Across these approaches, raw trajectories, derived state, runtime telemetry, and processing logic remain fragmented across files, databases, and scripts.

We argue that trajectories constitute a distinct data-management workload because several challenging properties appear together. First, trajectories have \textbf{nested and ordered execution structure}: a session contains turns, model invocations, tool calls, messages, and potentially branches, retries, and subagents. Second, their dominant payload is \textbf{free-form text}, so many important predicates require semantic evaluation rather than exact matching. Third, trajectories are \textbf{highly redundant}: successive model invocations often repeat much of the same context, causing large volumes of duplicated text to be stored and processed. Finally, trajectory processing creates \textbf{rich derived state and lineage}. A memory may depend on several messages, a failure judgment on a particular tool result and model response, and a training label on the context evaluated by a specific judge model and criterion. Efficiently managing such data therefore requires reasoning jointly over structure, order, semantics, redundancy, and provenance.

These properties create opportunities for cross-layer optimization that are difficult to realize when trajectories are split across log stores, vector databases, and task-specific pipelines. For example, a trajectory system should support queries such as: finding tool calls preceded by unsupported model claims; retrieving failures semantically similar to a given case together with the evidence used to classify them; or identifying model invocations that repeatedly process redundant context and correlating them with runtime cost. Such queries combine structural traversal, semantic predicates, temporal ordering, and provenance in a single analysis.


\vspace{0.8em}\noindent
\begin{minipage}{0.96\linewidth}
\vrule width 2pt\hspace{0.8em}%
\begin{minipage}{0.96\linewidth}
\textit{We therefore envision \textbf{TrajectoryDB}, a trajectory-native data management system that treats raw trajectories and their derived data as first-class citizens. \vspace{0.64em}}
\end{minipage}
\end{minipage}

TrajectoryDB provides a unified data model for execution records, their structural and temporal relationships, semantic annotations, and derived state, together with query operators that jointly reason over structure, order, semantics, and lineage. Rather than layering independent document, vector, graph, and log systems, TrajectoryDB co-designs ingestion, storage, and query processing around trajectory-specific access patterns and redundancies. Our goal is to make agent trajectories as natural to store, query, and optimize as documents are in document databases or time-series data are in time-series databases.

\section{Trajectory as a New Data Type}

\subsection{Agent Trajectory}


\noindent\textbf{Agent trajectory.} An agent can be viewed as a runtime loop that orchestrates interactions between a model and its environment. The basic unit of communication between an agent and the model is a \emph{\textbf{message}}, which contains \emph{a role and its associated content}. There are typically four roles: \texttt{system} refers to the system prompt, which specifies instructions that always apply, \texttt{user} provides user input, \texttt{assistant} records model responses, and \texttt{tool} captures results returned by external tools. To decide what to do next, the agent sends the model a request containing the current \emph{context}, i.e., the accumulated messages till now and other relevant states, and receives a response that it may then execute. We refer to one such model request and its resulting execution as a \emph{\textbf{step}}. A user input can trigger multiple steps before the agent returns a response to the user, which forms a \emph{\textbf{turn}}. A task often spans multiple turns, and the resulting sequence of turns forms a \emph{\textbf{session}}. 
We call the sequence of requests~\footnote{Each request carries an accumulative context.}, messages, and tool executions produced while completing a task as an \textbf{agent trajectory}. 
 

\noindent\textbf{Execution configuration.} Beyond the trajectory itself, agent execution also produces information often queried alongside it, including the configuration used to instantiate the agent and telemetry from the underlying system. A trajectory often also records the components and settings that shaped its execution, including the model, available tools and their versions, and sampling parameters such as temperature and random seed. Together, this metadata describes the environment in which the agent interacted and supports reproducibility, debugging, and comparison across runs.


\noindent\textbf{Operational telemetry.}
A trajectory may also include system-level measurements associated with its execution. Time to first token (TTFT) and per-step latency can support service-level agreement (SLA) and user-experience (UX) monitoring. Input and output token counts, together with cache-hit ratios, can inform cost and resource optimization. Retries, rate-limit events, timeouts, and error codes provide signals for diagnosing failures and triggering operational alerts. Tool call resource usage helps identify opportunities to improve agent execution.




\begin{figure}
  \centering
  \includegraphics[width=\linewidth]{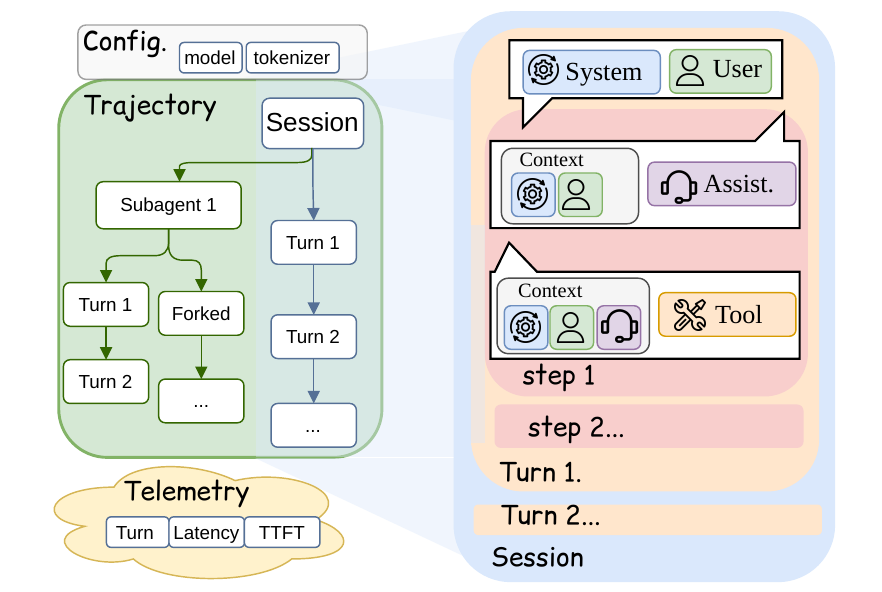}
  \caption{Agent trajectories can branch through subagents and forks, forming a tree of sessions. Within each session, execution follows a session-turn-step-message hierarchy. 
  Trajectories are not just logs; they are evolving, structured execution records with semantic content, repeated context, and derived data whose provenance matters. Existing systems each handle only part of this.
Alongside the trajectory, configuration metadata records the model and settings that produced it, while telemetry captures runtime performance.}
  \label{fig:agentic_flow}
\end{figure}

\subsection{Diverse Use Cases}
Trajectories support a range of downstream tasks. We group six representative use cases into three categories: \begin{enumerate*}[label=(\Alph*)] \item learning from past experience, \item analyzing and controlling agent behavior, and \item monitoring and optimizing runtime execution. \end{enumerate*}


\noindent\textbf{A.1 Extracting reusable memory.}
Agents frequently \emph{extract memory} from past trajectories~\cite{zhao2024expel,fu2024autoguide,ouyang2026reasoningbank,zhang2026ace}. These methods distill lessons from reasoning, tool calls, observations, and final task artifacts such as a markdown-formatted webpage. The resulting memory can guide future tasks without replaying the full trajectory.

\noindent\textbf{A.2 Fine-tuning from successful trajectories.}
Trajectories also serve as training data. In rejection-sampling fine-tuning (RFT)~\cite{yuan2023rft}, as used by SWE-Gym~\cite{pan2025swegym}, SWE-smith~\cite{yang2025swesmith}, and R2E-Gym~\cite{jain2025r2egym}, successful trajectories are rendered into the student model's chat template. By autoregressive optimization, successful multi-step behavior is transferred into the model.

\noindent\textbf{B.1 Root-cause analysis.}
Trajectories support failure diagnosis and anomaly detection. For example, 
TRAIL~\cite{deshpande2025trail} identifies reasoning, planning, and system-execution errors in the trajectory; 
AgentRx~\cite{barke2026agentrx} checks each step against constraints derived from tool schemas and domain policies, then uses the resulting violations to localize the first unrecoverable failure. 

\noindent\textbf{B.2 Enforcing safety and security.}
Trajectories help to reveal unsafe behavior. Tool-level checks inspect allowed actions, arguments, and anomalous call sequences~\cite{finos2026air,wiz2026guardrails}, while LLM monitors inspect an agent's reasoning for signs of test subversion or deception~\cite{baker2025cotmonitoring}. Importantly, a sequence can be unsafe even when each action appears acceptable in isolation. OpenAI observed this behavior in long-running agents and introduced monitoring over the entire evolving trajectory rather than individual actions~\cite{openai2026longhorizon}. 

\noindent\textbf{C.1 Monitoring runtime health.}
Operational telemetry supports performance and cost monitoring. Systems such as LangSmith, Langfuse, and Datadog record model and tool executions together with latency, token usage, cost, and errors~\cite{langsmithobservability,langfuseobservability,datadogobservability}. Persisting this telemetry enables both longitudinal analysis and post-hoc diagnosis: operators can detect regressions or cost spikes across runs and then trace an anomaly back to the model or tool call that caused it.

\noindent\textbf{C.2 System resource usage and scheduling optimizations.}
Agents frequently block on LLM inference and tool execution, making the scheduling order of LLM requests and tool calls a key determinant of end-to-end performance. Effective scheduling therefore requires exposing agent-level dependency and critical-path information to both LLM serving engines and operating-system schedulers, enabling them to prioritize latency-critical work and allocate resources accordingly~\cite{lin2024parrot,luo2026agentix,wang2026mars}.

\section{{Trajectory-DB} Requirements}
\label{sec:data_operations}
Despite the diverse use cases, these workloads repeatedly access, filter, and aggregate overlapping parts of a trajectory. We next summarize common data operations into three requirements that a database engine should support.

\subsection{Structure-aware Retrieval}

\subsubsection{Session-structure retrieval.}
Agent trajectories are naturally hierarchical: a session contains multiple turns, each turn contains multiple agent steps, and each step contains model requests and role-specific messages. Different workloads access this hierarchy at different granularities, from entire sessions to individual steps or messages. A trajectory database should therefore expose this structure explicitly, instead of treating each trajectory as an opaque blob, to support efficient fine-grained retrieval and processing.

\begin{table*}[t]
\centering
\caption{Mapping trajectory requirements to system challenges.}
\label{tab:requirements_mapping}

\begin{minipage}[t]{0.63\textwidth}
\vspace{0pt}
\centering

\renewcommand{\arraystretch}{1.2}
\setlength{\tabcolsep}{3pt}

\begin{tabularx}{\linewidth}{
  |>{\raggedright\arraybackslash}X
  |c|c|c|c|c|c|c|
}
\hline
\textbf{Requirements}
& \textbf{C1}
& \textbf{C2}
& \textbf{C3}
& \textbf{C4}
& \textbf{C5}
& \textbf{C6}
& \textbf{C7}
\\
\hline

3.1.1 Session-structure retrieval
& \checkmark
& \checkmark
& \checkmark
& \checkmark
& \checkmark
&
&
\\
\hline

3.1.2 Sequence retrieval
&
& 
& \checkmark
&
& \checkmark
& \checkmark
& 
\\
\hline

3.2.1 Composing semantic and relational operators
& 
& 
&
& \checkmark
& \checkmark
& \checkmark
& \checkmark
\\
\hline

3.2.2 Incremental semantic evaluation
& \checkmark
&
&
& \checkmark
&
&
& \checkmark
\\
\hline

3.3.1 Context reconstruction
& \checkmark
& \checkmark
&
& \checkmark
& \checkmark
&
&
\\
\hline

3.3.2 Preserving lineage for semantic judgments
&
& \checkmark
& \checkmark
& 
&
& \checkmark
& \checkmark
\\
\hline
\end{tabularx}
\end{minipage}
\hfill
\begin{minipage}[t]{0.36\textwidth}
\small
\vspace*{0.2\baselineskip}

\noindent\textbf{Challenges:}\par
\vspace{3pt}

\noindent
\textbf{C1:} Reconstructing request hierarchies.\par
\vspace{2pt}

\noindent
\textbf{C2:} Supporting graph structures over relational data.\par
\vspace{2pt}

\noindent
\textbf{C3:} Reconciling conflicting physical layouts.\par
\vspace{2pt}

\noindent
\textbf{C4:} Storing large text payloads efficiently.\par
\vspace{2pt}

\noindent
\textbf{C5:} Enabling order-sensitive query processing.\par
\vspace{2pt}

\noindent
\textbf{C6:} Co-optimizing relational and semantic operators.\par
\vspace{2pt}

\noindent
\textbf{C7:} Caching semantic judgments.
\end{minipage}

\end{table*}

\noindent\textbf{Session retrieval. }
Some workloads consume complete \emph{sessions}. For example, ExpeL~\cite{zhao2024expel} and AutoGuide~\cite{fu2024autoguide} compare successful and failed sessions on similar tasks to identify reusable lessons. A session, however, is not always a linear sequence when an agent spawns subagents, as shown in~\autoref{fig:agentic_flow}. Each subagent creates a branch rooted at the step that spawned it, so session retrieval must preserve this parent-child structure. 



\noindent\textbf{Step retrieval.}
Failure diagnosis often operates at the granularity of a step. For example, AgentRx~\cite{barke2026agentrx} derives execution rules from policies and tool definitions, then retrieves each step and checks it against those rules.
Some checks depend on multiple steps. For example, detecting whether an agent asks again for information it already obtained requires comparing the current step with prior tool results in the session.





\noindent\textbf{Role retrieval.}
Fine-tuning has to distinguish the messages based on \emph{roles}. The full conversation is used as input for training, but the loss is typically computed only over tokens generated by the assistant. The roles must therefore be stored as an explicit attribute so that only assistant messages are selected efficiently. 



\subsubsection{Sequence retrieval.}
Many trajectory workloads identify patterns over ordered events rather than individual records. Memory extraction, for example, identifies recurring \emph{tool-use sequences} and turns them into reusable lessons~\cite{fu2024autoguide,ouyang2026reasoningbank}: \emph{edit $\rightarrow$ failed test $\rightarrow$ edit the same file} may indicate an ineffective strategy, while \emph{search $\rightarrow$ read $\rightarrow$ edit $\rightarrow$ passing test} suggests a successful one. Sequence patterns can also span different message roles and actions. A safety policy may, for example, require \emph{assistant describes action $\rightarrow$ user confirms $\rightarrow$ action}, making the presence and order of the confirmation part of the check. Operational analysis introduces yet another \emph{event sequence}: LLM requests, retries, tool executions, and their telemetry can be matched in sequence to identify recovery behavior, repeated failures, or latency-critical paths. Sequence detection must therefore operate over heterogeneous trajectory events while preserving their execution order.
\subsection{LLMs as UDFs}
Trajectory analysis often requires predicates over the meaning of free-form text rather than exact matches over literals. Exposing LLM-based analysis as a general UDF allows the same semantic primitive to be reused across workloads and composed with other query operators.

\subsubsection{Composing semantic and relational operators.}
Semantic evaluation in trajectory queries is often embedded within a larger relational plan: its inputs are assembled from multiple records, and its output may feed downstream joins, ordering, or aggregation~\cite{patel2025semanticoptimization,mang2026plop}. For example, consider a root-cause query that asks for \emph{the first step whose tool call violates a constraint}, as in step-level failure diagnosis~\cite{barke2026agentrx}. The engine must group records by session, assemble the context needed to judge each step, apply the semantic predicate, and then use trajectory order to select the first violating step. Other queries compose semantic judgments with telemetry. To determine whether latency spikes are associated with recovery attempts after tool failures, for example, the engine first uses semantic evaluation to identify failures and then joins those judgments with runtime measurements~\cite{langsmithobservability,langfuseobservability,datadogobservability}. In these cases, semantic operators need to compose with relational and order-sensitive operators throughout the query plan.

\subsubsection{Incremental semantic evaluation over growing trajectories.}
Semantic predicates often operate on growing inputs, so their results must be updated statefully to remain valid as new evidence arrives. For example, as new sessions apply a memory, each outcome provides new evidence about whether that memory remains valid. The judge updates its prior assessment with the new outcome, and accumulated conflicts may trigger revision or invalidation of the memory. Safety monitoring follows the same pattern: as new actions arrive, the judge updates its assessment of whether the agent is drifting toward unsafe or off-plan behavior. Runtime-health analysis also compares a newly observed failure with recent failures to determine whether they share the same underlying fault. In each case, semantic evaluation must build on previous state and incorporate new evidence as the trajectory grows.

\subsection{Lineage Query}

\subsubsection{Context reconstruction.}
Long-running agents often compact their context by replacing earlier messages with summaries~\cite{anthropic2025contextengineering,li2026compactionrl}. After compaction, a model request contains summaries together with recent messages rather than the full original history. Several workloads still need to recover that history. A failure mode in root-cause analysis is that an agent violates a constraint because the relevant message was summarized away. Diagnosing such failure requires reconstructing the pre-compaction context to identify which compaction removed the information needed. Memory verification similarly requires the original trajectory to locate the evidence to support the memory~\cite{ouyang2026reasoningbank}.
Evaluating a different compaction strategy also requires the original history so that an alternative compacted context can be constructed and replayed~\cite{li2026compactionrl}.




\subsubsection{Preserving lineage for semantic operator.}
TrajectoryDB should preserve not only each semantic operator result, but also its provenance: the evaluated input, supporting trajectory records, judge model, evaluation criterion, and other information needed to reconstruct how the result was derived.
For fine-tuning, an LLM judge may label whether a message is supported by its context, and the resulting label can be reused across subsequent training runs. Other use cases require finer-grained lineage. For example, linking each extracted memory or failure judgment to the specific trajectory evidence that supports it~\cite{fu2024autoguide,ouyang2026reasoningbank,barke2026agentrx}.
This lineage serves two purposes. First, it enables invalidation: if the source trajectory changes, the memory becomes stale, or the judge model or criterion is updated, the system can identify which judgments need to be reevaluated. Second, it supports provenance queries that ask what produced a judgment and which trajectory evidence supports it. Semantic judgments therefore become reusable derived data rather than outputs of one-time LLM calls.

\section{Challenges and Opportunities}
In this section, we map the requirements above to the key challenges as shown in~\autoref{tab:requirements_mapping}, and discuss where existing database systems fall short. We collect a one-month trace from our production system ~\cite{freeinference} and use it to characterize agent trajectory behavior in this section.

\subsection{Structure Reconstruction and Maintenance}
Trajectory structure is scattered across timestamped requests. Therefore, TrajectoryDB should reconstruct this structure and organize it both logically and physically.

\subsubsection{Reconstructing Request Hierarchy}
\mybox{Challenge: Missing structural links.}
When agent activity is stored as generic records, relationships among sessions, turns, steps, and messages are often implicit. Without session identifiers, requests may need to be associated through shared context. Turn and step boundaries may likewise need to be inferred from events such as user messages and completed tool calls. Recovering the execution hierarchy therefore requires nontrivial reconstruction.

\noindent\textbf{Design question: Which structures should be reconstructed eagerly?}
Eager reconstruction simplifies downstream queries but adds work to the ingestion path. OpenTelemetry notes that complex filtering and transformation can significantly reduce Collector performance~\cite{oteltransform}; ClickHouse similarly found that parsing and transformation in its OTel pipeline consumed substantial CPU, and replacing this path enabled $20\times$ higher throughput with less than $10\%$ of the previous CPU footprint~\cite{clickhouse100pb}. Conversely, retaining only timestamped records shifts reconstruction cost to every query. TrajectoryDB must therefore determine which structural relationships to materialize during ingestion and which to derive lazily.



\mybox{Challenge: Requests can be separated by long gaps.}
Requests belonging to the same long-running agent trajectory may arrive far apart in time. In fact, consecutive requests from the same workflow can be separated by hours or days, as shown in~\autoref{fig:freeinference}. Moreover, while most sessions are short-lived, the tail can last weeks. The ingestion layer must therefore associate each new request with a trajectory state that may have been created much earlier.

\noindent\textbf{Design question: How to maintain state for long-lived streaming trajectories?}
To correlate each incoming request with earlier requests from the same session, the ingestion layer must retain per-session state, such as accumulated context and trajectory metadata. Stateful stream-processing systems~\cite{flinklateness,kafkastreamsgrace,sparkwatermark} support such cross-record correlation, but typically rely on a notion of when retained state can safely be finalized or discarded, using mechanisms such as watermarks or grace periods. Agent trajectories provide no natural cutoff: a seemingly inactive session may resume hours or even days later. Setting a short inactivity bound risks prematurely splitting a single trajectory, whereas setting a long bound leaves many inactive sessions open and increases state overhead. Existing stream-processing abstractions therefore do not naturally support incremental reconstruction of long-lived agent trajectories without assuming a bounded or predictable completion window.

\subsubsection{Maintaining Graph Structure over Relational Data}

\mybox{Challenge: Trajectory relationships form graphs, not flat rows.}
Agent trajectories contain graph-structured relationships. Subagent execution creates tree-shaped branches, while lineage introduces labeled many-to-many edges: memories are \emph{derived from} evidence, judgments are \emph{supported by} or \emph{inconsistent with} earlier records, and failures may be \emph{caused by} or \emph{repeat} earlier failures. Together, these relationships form DAGs over trajectory records.

\noindent\textbf{Design question: How to efficiently support both graph traversal and relational analytics?}
Relational engines can encode edges using fields such as \texttt{parent\_id}, but traversal then requires recursive queries such as PostgreSQL recursive CTEs~\cite{postgresqlrecursive}. 
These queries can be expensive for deep or branching graphs and do not inherently preserve traversal order~\cite{postgresqlplanner,postgresqlrecursive}. Downstream sequence or window analytics must therefore reconstruct that order explicitly~\cite{postgresqlwindow}.
Native graph engines support traversal directly, but storing large trajectory payloads can increase I/O and deserialization cost. Separating graph topology from relational payloads avoids this overhead but introduces costly cross-system coordination.
SQL/PGQ integrates graph patterns with SQL~\cite{deutsch2021graph}, while Chimera~\cite{lee2025chimera} combines relational storage with native adjacency structures. However, traversal may still require joins or pointer chasing between topology and payloads. Agent trajectories are more specialized: they are append-only, mostly forest-shaped, and naturally clustered by session. Exploiting these properties may enable tighter integration of graph traversal with relational and ordered access.


\subsection{Data Layout and Storage}
\subsubsection{Reconciling Conflicting Physical Layouts}
\mybox{Challenge: Use cases favor conflicting physical layouts.}
Trajectory workloads impose different locality requirements on the same data. Replay and context reconstruction scan full requests in session order, while sequence matching searches ordered requests for particular subsequences; both favor \textit{session-contiguous layouts}. In contrast, role filtering, tool-pattern mining, and telemetry analysis access only a few fields across many sessions, favoring \textit{columnar layouts}. Lineage queries introduce a third pattern, where \textit{locality follows graph links} among records rather than their session or field organization.

\noindent\textbf{Design question: How should trajectories be physically organized for diverse access patterns?}
No single layout naturally serves all three. Transactional databases such as PostgreSQL are attractive primary stores because trajectory recording often occurs alongside application state such as users, authentication, and billing. However, large analytical scans over growing trajectory volumes can become expensive. 

TrajectoryDB must therefore determine the primary representation, which auxiliary layouts or indexes to maintain, and how to keep them consistent as trajectories continuously grow.

\subsubsection{Storing Large but Redundant Text Payload}
\begin{figure}[t]
\centering
\begin{subfigure}[t]{0.49\linewidth}
    \centering
    \includegraphics[width=\linewidth]{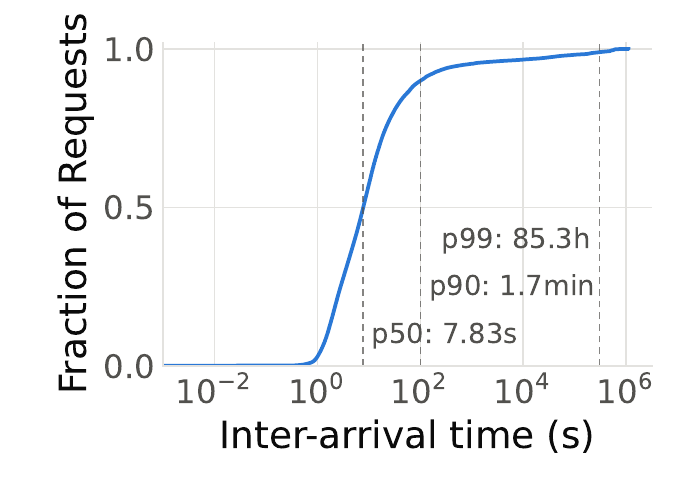}
    \caption{}
    \label{fig:gap}
\end{subfigure}\hfill%
\begin{subfigure}[t]{0.48\linewidth}
    \centering
    \includegraphics[width=\linewidth]{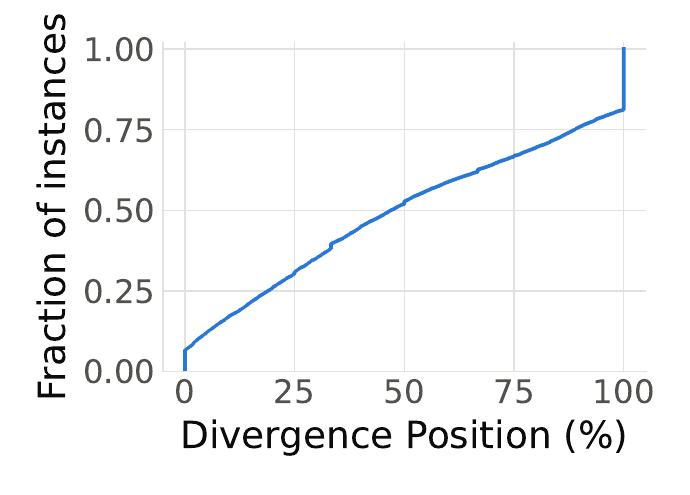}
    \caption{}
    \label{fig:diverge}
\end{subfigure}
\caption{(a) CDF of request inter-arrival time (IAT); (b) 80\% of the sessions mutate the prefix during lifetime, and the figure shows a CDF of divergence position, normalized by the total session length. Data from production\protect\footnotemark.}
\label{fig:freeinference}
\vspace{-2em}
\end{figure}

\footnotetext{We operate https://freeinference.org, which serves half a trillion tokens every month.}
\mybox{Challenge: Trajectories are requests with cumulative context.}
Each turn sends the full context with all the previous turns: the $n^{th}$ request in the trajectory contains all the messages in the previous $n-1$ turns. 
Continuously recording agent inputs and outputs can generate enormous amounts of text. Assuming a \textit{conservative} throughput of 2,000 output tokens/s per H100~\cite{nvidiah100inference} and a 50:1 input-to-output ratio~\cite{togethercoding2026}, a 1 GW AI facility with 1,000,000 fully utilized H100s would generate roughly 102 billion logged tokens/s. At approximately four bytes per token~\cite{openaitokens}, naively storing all input and output tokens would require about 35 PB/day.
Such large amount of text becomes the payload of the prompt column if storing the request in a table.

\noindent\textbf{Design question: Should the text payload be stored inline or separately?}
Large payloads are often stored outside of the table. PostgreSQL, for example, splits oversized values into chunks in a separate TOAST table and accesses them through pointers~\cite{postgresqltoast}. This avoids filling the main table with large messages, but breaks the contiguous layout preferred by session-oriented scans. Keeping the payloads inline has the opposite problem: a page may hold only a few messages, causing unnecessary I/O and cache pollution. TrajectoryDB therefore needs a layout that keeps large text efficient to scan without letting it dominate the record storage.

\mybox{Challenge: Context is highly redundant, but not always prefix-aligned.}
Each request typically carries all the messages from the same session as accumulated context, often referred to as a \emph{prefix}, so storing requests independently duplicates the same prefix many times.
The challenge is that context sharing is not always a clean prefix. 
In production, we observe that \textit{80\% of the sessions have context mutation: compaction, truncation of tool results, or edits to earlier context break prefix sharing. This makes it impossible to store only the newly appended turn. }
For clean sessions, \autoref{fig:diverge} shows where sessions diverge, with each fork creating a new instance. A sample of 5,951 sessions from our production systems, 3,794 never fork and therefore contribute only one instance each. The remaining sessions fork into multiple instances and account for over 80\% of all instances. These forks occur throughout the sessions without concentrating at a particular stage, while non-forking sessions account for less than 20\% of all instances.

\noindent\textbf{Design question: How to support prefix-aware text compression and deduplication?}
Existing compression is unaware of request-level sharing.
Database compression usually works within a page or block. If the same message appears in requests stored in different blocks, that redundancy is not shared~\cite{oraclerowcompression,innodbpagecompression,clickhousecompression}.
Dictionary encoding reuses identical values across rows, but trajectory payloads are long and mostly distinct. The dictionary therefore grows to the size of the data itself without exploiting redundancy across different chunks~\cite{parquetencoding}. Data exported from our production PostgreSQL instance achieves a 20$\times$ compression ratio with Zstandard, compared with less than 2$\times$ using PostgreSQL’s built-in compression. 
For trajectories, a new request extends an earlier context until compaction, truncation, or another update causes the histories to diverge. The storage engine should exploit this known sharing directly rather than rediscovering redundancy from raw bytes or literals within physical data blocks.
\subsection{Query Language and Optimization}

\subsubsection{Enabling Order-Sensitive Query Processing}

\mybox{Challenge: Trajectory queries are often order-sensitive.}
For trajectory queries, order can determine the answer rather than merely its presentation. Sequence matching, for example, must distinguish \emph{edit $\rightarrow$ failed test $\rightarrow$ edit} from \emph{edit $\rightarrow$ edit $\rightarrow$ failed test}, despite containing the same events. Likewise, a rolling reward delta is meaningful only when records are partitioned by session and ordered by request arrival~\cite{postgresqlwindow}.
The relevant order depends on query scope. Within a subagent branch, events follow a natural execution order. Across subagents, however, branches may execute concurrently, so no single total order exists. A trajectory query model must therefore support both branch-local order and partial order across branches.

\noindent\textbf{Design question: How to preserve order and optimize the query plan?}
Current engines treat order as a local property rather than a first-class property of the query plan, while some document stores provide limited native support for ordered analytics~\cite{mongodb,documentdb}. Indexes can produce records in the desired order and avoid explicit sorting~\cite{postgresqlindexorder}, but subsequent joins, aggregation, or parallel execution may destroy that order. A trajectory-aware engine should therefore track order through operators, reuse existing order rather than recompute it, propagate it through lineage traversal, and explicitly represent partial order across concurrent branches.

\subsubsection{Co-optimization Across Relational and Semantic Operators}

\mybox{Challenge: Semantic operators consume and produce relational data.}
Trajectory queries often require semantic evaluation because key information is embedded in free-form text, while the resulting judgments are further processed by relational operators. The optimizer must therefore choose an execution order across conventional operators and expensive LLM calls. Pushing a cheap telemetry filter before a semantic judge can reduce LLM invocations, whereas an early, highly selective semantic filter may avoid an expensive traversal. Choosing between these plans requires estimating both the cost and selectivity of relational and semantic operators.

\noindent\textbf{Design question: How to define a declarative framework to
integrate semantic operators into the relational plan?}
Existing semantic-query systems typically fix the input shape of each LLM operator. LOTUS, for example, defines semantic filters over individual tuples, joins over tuple pairs, and aggregations over a declared relation~\cite{patel2025semanticoptimization}. Trajectory queries may instead require an LLM to consume dynamically constructed inputs---such as an ordered subsequence, a lineage neighborhood, or accumulated session context---and then feed its judgment into subsequent relational processing. Supporting such queries requires semantic operators that compose naturally with traversal, ordering, filtering, and aggregation, while remaining visible to the optimizer for joint planning.


\subsubsection{Caching Semantic Results}
\mybox{Challenge: Semantic results are expensive and ephemeral.}
Semantic evaluation is expensive, but can be reused between queries. A memory verifier, safety monitor, and failure-analysis query may ask the same question of whether a session fails. Caching intermediate semantic results avoids such repeated LLM calls. However, these results are fragile: changes to the query, trajectory state, judge model, or evaluation criterion may invalidate them.

\noindent\textbf{Design question: When can a cached semantic result be safely reused?}
Existing semantic caches such as GPTCache~\cite{fu2023gptcache} and MeanCache~\cite{gill2025meancache} reuse LLM outputs by matching semantically similar inputs. This is insufficient for trajectory queries, where validity depends not only on input similarity but also on the underlying trajectory state. As new steps arrive, previously cached judgments may become stale. Semantic caching for trajectories must therefore be \emph{lineage-aware}: each cached result should record the trajectory state, query semantics, and model configuration on which it depends, so the system can identify and refresh only results affected by subsequent changes.



\section{Related Work}

\noindent\textbf{Semantic operators.} LOTUS~\cite{patel2025semanticoptimization} and Palimpzest~\cite{liu2025palimpzest} provide declarative frameworks with LLM-powered operators over relational data. Building on Palimpzest's optimizations for filters and joins, Abacus~\cite{russo2026abacus} formalizes cost-based optimization for semantic operator systems, searching over model choice, prompting strategy, and cost--quality trade-offs. PLOP~\cite{mang2026plop} further optimizes the placement of semantic operators, while UQE~\cite{dai2024uqe} uses a small model and classifier to identify candidate rows for a semantic predicate and avoid invoking an LLM on every row. Semantic operators are also used for data discovery and preprocessing. DocETL~\cite{shankar2025docetl} uses LLM-powered operators to discover schemas and structure documents into tables, while EVAPORATE~\cite{arora2023evaporate} uses LLMs to generate extraction code for document processing.

\noindent\textbf{Data lineage.} Data lineage traces which input tuples contribute to each output tuple, supporting view maintenance and source-level explanation~\cite{cui2000lineage}. Titian~\cite{interlandi2015titian} captures record-level lineage in distributed dataflow jobs. Smoke~\cite{psallidas2018smoke} co-designs lineage capture with physical operators to achieve low capture overhead and interactive lineage queries. In contrast, GProM~\cite{arab2018gprom} avoids materializing lineage by rewriting queries to compute provenance on demand, trading query-time cost for zero capture overhead.

\section{Conclusion}
Agent trajectories are emerging as a distinct data type. They combine hierarchical execution structure, large volumes of text, and complex lineage relationships among trajectory records and derived state. Existing database systems are not designed to support these properties together. We therefore envision TrajectoryDB that co-designs ingestion, storage, and query processing to optimize the full trajectory data lifecycle.


\bibliographystyle{ACM-Reference-Format}
\bibliography{sample}

\end{document}